\documentclass[conference]{IEEEtran}
\IEEEoverridecommandlockouts
\usepackage{cite}
\usepackage{amsmath,amssymb,amsfonts}
\usepackage{algorithmic}
\usepackage{graphicx}
\usepackage{textcomp}
\usepackage{xcolor}
\def\BibTeX{{\rm B\kern-.05em{\sc i\kern-.025em b}\kern-.08em
    T\kern-.1667em\lower.7ex\hbox{E}\kern-.125emX}}

\usepackage{amsmath,amssymb}
\usepackage{extarrows}
\usepackage{mathtools}
\usepackage{nicefrac}
\usepackage{mathrsfs}
\usepackage{bm}
\usepackage{tikz}
\usetikzlibrary{arrows,decorations.pathmorphing}    
\usepackage{circuitikz}
\usetikzlibrary{circuits.ee.IEC}
\usetikzlibrary{dsp,chains,backgrounds}
\usetikzlibrary{shapes.arrows, fadings,shapes.geometric}
\usetikzlibrary{patterns}
\usepackage{bm}
\usepackage{siunitx} 
\usepackage{pgfplots} 							
\pgfplotsset{compat=newest}						
\pgfplotsset{plot coordinates/math parser=false} 		

\newcommand{\Kprim}{\bm{K}}
\newcommand{\Kadj}{\tilde{\bm{K}}}
\newcommand{\As}{\bm{\mathcal{A}}}
\newcommand{\Bs}{\bm{\mathcal{B}}}
\newcommand{\Cs}{\bm{\mathcal{C}}}
\newcommand{\Ks}{\bm{\mathcal{K}}}

\newcommand{\her}{{\text{\tiny$\mathrm{H}$}}}
\newcommand{\tra}{{\text{\tiny$\mathrm{T}$}}}

\newcommand{\expE}[1]{\mathrm{e}^{#1}}

\newcommand{\Lp}{\bm{\mathrm{L}}}

\newcommand{\linpr}{\left\langle}
\newcommand{\rinpr}{\right\rangle}

\renewcommand{\baselinestretch}{0.94}

\begin{document}

\title{Spherical Diffusion Model with Semi-Permeable Boundary: A Transfer Function Approach 
}

\author{\IEEEauthorblockN{Maximilian Sch\"{a}fer\textsuperscript{*}, Wayan Wicke\textsuperscript{*}, Werner Haselmayr\textsuperscript{\textdagger}, Rudolf Rabenstein\textsuperscript{*}, and Robert Schober\textsuperscript{*}}
\IEEEauthorblockA{\textit{\textsuperscript{*}Friedrich-Alexander University Erlangen-N\"urnberg (FAU),} Erlangen, Germany 
\\
	\textit{\textsuperscript{\textdagger}Johannes Kepler University (JKU),}
	Linz, Austria}
}
\maketitle

\begin{abstract}
The derivation of suitable analytical models is an important step for the design and analysis of molecular communication systems.
However, many existing models have limited applicability in practical scenarios due to various simplifications (e.g., assumption of an unbounded environment). In this paper, we develop a realistic model for particle diffusion in a bounded sphere and particle transport through a semi-permeable boundary. 
This model can be used for various applications, such as modeling of inter-/intra-cell communication or the release process of drug carriers. The proposed analytical model is based on a transfer function approach, which allows for fast numerical evaluation and provides insights into the impact of the relevant molecular communication system parameters. The proposed solution of the bounded spherical diffusion problem is formulated in terms of a state-space description and the semi-permeable boundary is accounted for by a feedback loop. Particle-based simulations verify 
the proposed modeling approach. 


\end{abstract}


\section{Introduction}
\label{sec:intro}
Molecular communications (MC) is a promising approach for environments and applications, where conventional wireless communications using electromagnetic waves is not feasible or detrimental, such as for smart drug delivery inside the human body~\cite{7405285,Chude-Okonkwo_et_al:IEEECommSurvTutorials:2017}. 
For the analysis of natural and for the design of synthetic MC systems, the development of suitable models for the individual components of the MC system is a crucial step. 
In recent years, various models for the transmitter~(TX), propagation channel, and  receiver (RX) have been developed~\cite{jamali_channel_2018}. However, many existing models do not properly reflect the characteristics of practical scenarios, such 
as bounded environments and the properties of the TX.

In particular, in most existing MC works, the~TX is assumed to be an ideal point source that can produce the molecules instantaneously which then immediately enter the physical channel~\cite{7405285}. Hence, this simplified model neglects the effect of the~TX geometry, the molecule generation process, and the molecular release mechanism~\cite{jamali_channel_2018}. However, there have been several works that partially take these effects into account, which leads to more realistic models. A spherical~TX with molecules distributed over a virtual sphere (i.e., a transparent surface) has been proposed in~\cite{noel:comm:2016}, which corresponds to a more realistic model since in reality molecules occupy space. Moreover, a spherical~TX with the molecules distributed over a reflective surface is considered in~\cite{noel:comm:2016,Yilmaz_17}.\footnote{Please note that \cite{noel:comm:2016,Yilmaz_17} only provide particle-based simulations for this scenario.} Although this model considers the effect of the~TX geometry, it neglects the particle generation and the release mechanism. 
In~\cite{Chude:2015}, the~TX is modeled by a sphere whose surface is covered by nanopores and a point source for the molecular production at its center. 
The proposed model only considers the effect of the TX surface on the released molecules via a 
modified diffusion coefficient.
In~\cite{Arjmandi:ieee:2016}, the TX is modeled 
as a spherical object with ion channels in its membrane. 
The molecule release is controlled via opening and closing of the ion channels, which in turn is controlled by a voltage across the membrane. 
Based on the TX geometry, the mechanism controlling the release, and the supposed particle generation process, a closed-form expression for the modulated signal is derived. 
In this work, we develop an analytical model for particle diffusion in a bounded sphere and particle transport through a space and time-variant semi-permeable boundary. This model can serve as a realistic TX model, which uses the permeability of the boundary to control the dynamics of the particle release. Moreover, the presented model can be used to describe the impact of cell aging on the membrane permeability~\cite{Yury_95} and to model monolithic controlled drug release~\mbox{\cite{giret:chem:2015, lee:pharm:2011}}. The proposed model is based on a formulation in terms of transfer functions, which also allows to incorporate arbitrary particle generation models. The diffusion problem in a bounded sphere with semi-permeable boundary was recently considered in \cite{Arjmandi:ieee:2019} using a Green's function approach. However, in contrast to \cite{Arjmandi:ieee:2019}, the proposed model allows to analyze space and time-variant permeability at the boundary and to model the interconnection of multiple spheres. Here, two spherical objects are interconnected by a semi-permeable membrane and delay-free ion channels.
By further extensions, the proposed approach can be applied to model the propagation of~Ca\textsuperscript{$2+$} waves through cell chains, i.e.,~Ca\textsuperscript{$2+$} signaling~\mbox{\cite{Barros:ieee:2015,Heren:IEEE_ICC-6649341:2013}}. This would lead to  
a more detailed characterization of these intercellular signaling dynamics compared to the existing one-dimensional closed form solutions~\cite{Harris_Timofeeva:PhysRevE.82.051910:2010}.

The main contributions of this paper can be summarized as follows:
\begin{itemize}
  \item We derive an analytical model for particle diffusion in a sphere with general boundary behavior based on the modal expansion of an initial-boundary value problem~(IBVP), which leads to a solution of the IBVP in terms of transfer functions~\mbox{\cite{churchill:1972,Zwart:SCL:2004,rabenstein:ijc:2017}}. The proposed model facilitates fast numerical evaluation and insightful closed-form solutions~(cf.~\cite{schaefer:icc:2019}). The model is formulated in terms of a state-space description (SSD). 
  \item This general model is extended to include a semi-permeable boundary. The semi-permeability is accounted for by a feedback loop~\cite{schaefer:molcom:2019a,rabenstein:ijc:2017}. Moreover, we further modify this model to allow for time-variant permeability.
  \item Based on the semi-permeable boundary, we model the interconnection of two spherical objects through a semi-permeable membrane and delay-free ion channels. These investigations can be easily extended to a larger number of interconnected spheres or more complex topologies.
  \item We numerically evaluate the proposed models and verify them by particle-based simulations, using the AcCoRD simulator~\cite{NOEL201744}. We observe that the proposed model significantly reduces the time for performance evaluation compared to particle-based simulations.
\end{itemize}
  
The remainder of this paper is organized as follows.
In Section~\ref{sec:physic}, we formulate an IBVP for particle diffusion in a bounded sphere. In Section~\ref{sec:model}, we derive a transfer function model for the diffusion process with general boundary conditions, which is formulated in terms of an SSD. Based on this general model, in Section~\ref{sec:semi}, we derive a model for a (time-variant) semi-permeable boundary that is accounted for by a feedback loop. 
Furthermore, in Section~\ref{sec:inter}, we investigate the interconnection of two spheres by a semi-permeable membrane and delay-free ion channels. Finally, we draw some conclusions in Section~\ref{sec:conc}.

\section{Physical Model}
\label{sec:physic}
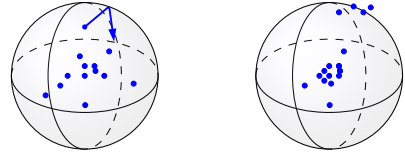
\begin{figure}[t]
	\centering
	\begin{tikzpicture}[auto, scale = 0.65, every node/.style={scale=0.65}, font=\LARGE, node distance=2.5cm,>=latex', 
	rounded corners=1pt,domain=0:10]
	
	\draw (-1.5,2.5) arc (180:360:1.5cm and 0.75cm);
	\draw[dashed] (-1.5,2.5) arc (180:0:1.5cm and 0.75cm);
	\draw (0,4) arc (90:270:0.75cm and 1.5cm);
	\draw[dashed] (0,4) arc (90:-90:0.75cm and 1.5cm);
	\draw (0,2.5) circle (1.5cm);
	\shade[ball color=blue!10!white,opacity=0.10] (0,2.5) circle (1.5cm);
	
%
	\draw[fill=blue, draw=blue] (0,2.5) circle (0.05cm);
	\draw[fill=blue, draw=blue] (0,2.7) circle (0.05cm);
	\draw[fill=blue, draw=blue] (0.4,2.5) circle (0.05cm);
	\draw[fill=blue, draw=blue] (-0.35,2.5) circle (0.05cm);
	\draw[fill=blue, draw=blue] (-0.1,2.9) circle (0.05cm);
	\draw[fill=blue, draw=blue] (1,2.34) circle (0.05cm);
	\draw[fill=blue, draw=blue] (-0.8,2.1) circle (0.05cm);
	\draw[fill=blue, draw=blue] (0.2,2.7) circle (0.05cm);
	\draw[fill=blue, draw=blue] (0.22,2.6) circle (0.05cm);
	\draw[fill=blue, draw=blue] (0.01,1.9) circle (0.05cm);
	\draw[fill=blue, draw=blue] (-0.5,2.3) circle (0.05cm);
		\draw[fill=blue, draw=blue] (0.5,3) circle (0.05cm);
	
	\draw[dspconn,blue](0,3.5) node[thick,circle,fill=blue,inner sep=0pt,minimum size=3pt]{} -- (0.5,3.92) -- (0.6,3.2);

	\draw (3.5,2.5) arc (180:360:1.5cm and 0.75cm);
	\draw[dashed] (3.5,2.5) arc (180:0:1.5cm and 0.75cm);
	\draw (5,4) arc (90:270:0.75cm and 1.5cm);
	\draw[dashed] (5,4) arc (90:-90:0.75cm and 1.5cm);
	\draw (5,2.5) circle (1.5cm);
	\shade[ball color=blue!10!white,opacity=0.10] (5,2.5) circle (1.5cm);
	
	\draw[fill=blue, draw=blue] (5,2.5) circle (0.05cm);
	\draw[fill=blue, draw=blue] (5,2.7) circle (0.05cm);
	\draw[fill=blue, draw=blue] (5.2,2.5) circle (0.05cm);
	\draw[fill=blue, draw=blue] (4.8,2.5) circle (0.05cm);
	\draw[fill=blue, draw=blue] (4.9,2.6) circle (0.05cm);
	\draw[fill=blue, draw=blue] (5.04,2.34) circle (0.05cm);
	\draw[fill=blue, draw=blue] (4.9,2.4) circle (0.05cm);
	\draw[fill=blue, draw=blue] (5.2,2.7) circle (0.05cm);
	\draw[fill=blue, draw=blue] (5.22,2.6) circle (0.05cm);
	\draw[fill=blue, draw=blue] (5.2,2.7) circle (0.05cm);
	\draw[fill=blue, draw=blue] (5.22,2.6) circle (0.05cm);
	\draw[fill=blue, draw=blue] (5.01,1.9) circle (0.05cm);
	\draw[fill=blue, draw=blue] (4.5,2.3) circle (0.05cm);
	\draw[fill=blue, draw=blue] (5.3,3) circle (0.05cm);

	\draw[fill=blue, draw=blue] (5.5,3.92) circle (0.05cm);
	\draw[fill=blue, draw=blue] (5.2,3.8) circle (0.05cm);
	\draw[fill=blue, draw=blue] (5.7,3.8) circle (0.05cm);
	\draw[fill=blue, draw=blue] (5.85,3.9) circle (0.05cm);
	
	\end{tikzpicture}
	\caption{Illustration of two bounded spheres with radius $R_0$, where the particles (blue) diffuse with a constant diffusion coefficient $D$. Left: The boundary of the sphere is fully reflective. Right: The boundary of the sphere is permeable.
	}
	\label{fig:1}
\end{figure}

The diffusion of particles (blue dots) in a bounded sphere with radius $R_0$ is shown in Fig.~\ref{fig:1}. Their movement in the sphere can be described by the well known diffusion equation based on Fick's laws in spherical coordinates \cite{crank:1975}. Depending on the boundary conditions, the particles are either reflected when they hit the boundary (Fig.~\ref{fig:1}, left hand side) or they can penetrate the boundary to leave the sphere (Fig.~\ref{fig:1}, right hand side). 
In this section, the diffusion of particles in a sphere is modeled by an IBVP. To allow for considering both transparent and reflective boundaries, 
a general set of boundary conditions is introduced. These conditions will be further specialized in Section~\ref{sec:semi} to support permeable boundaries.
%

\subsection{Initial-boundary Value Problem}
\label{subsec:ibvp}

The diffusion in a three-dimensional (3D) sphere is described in spherical coordinates by the dynamics of particle concentration $p(\bm{x},t)$ and 3D flux vector $\bm{i}(\bm{x},t)$ 
which are functions of 
time $t$ and 3D space coordinate $\bm{x} = \left[r, \varphi, \theta\right]^\tra$, where $(\cdot)^\tra$ denotes transposition.
Defining the Volume $V$ of the sphere and its boundary $\partial V$ as
\begin{align}
\begin{split}
V &\coloneqq \{\bm{x}\!: \left[r, \varphi, \theta\right]^\tra \big\vert r \in [0,R_0], \varphi \in [-\pi, \pi], \theta\in [0, \pi] \},\\
\partial V &\coloneqq \{\bm{x}\!: \left[r, \varphi, \theta\right]^\tra \big\vert r = R_0, \varphi \in [-\pi, \pi], \theta\in [0, \pi] \},
\end{split}
\label{eq:0}
\end{align}
spherical diffusion can be described in terms of partial differential equations (PDEs) as follows \cite{crank:1975}
\begin{align}
\bm{i}(\bm{x},t) + {D}\,\bm{\mathrm{grad}}\,p(\bm{x},t) &= 0,\label{eq:1}\\
\frac{\partial}{\partial t} p(\bm{x},t) + \bm{\mathrm{div}}\,\bm{i}(\bm{x},t) &= f_\mathrm{e}(\bm{x},t),\label{eq:2}
\end{align}
with the $3\times 1$ vector 
\begin{align}
\bm{i}(\bm{x},t) = \begin{bmatrix}
i_r(\bm{x},t) & i_\varphi(\bm{x},t) & i_\theta(\bm{x},t)
\end{bmatrix}^\tra . \label{eq:4}
\end{align}
The operators $\bm{\mathrm{grad}}$ and $\bm{\mathrm{div}}$ in \eqref{eq:1} and \eqref{eq:2} denote the gradient and divergence in spherical coordinates, respectively. Time derivatives are denoted by $\frac{\partial}{\partial t}$. The particles diffuse with a constant isotropic diffusion coefficient $D$. The function $f_\mathrm{e}$ in \eqref{eq:2} is a source term that allows to model the spatially and temporally distributed release of particles in the sphere. 

At the boundary of the sphere, a Dirichlet boundary condition for the flux in radial direction is defined
\begin{align}
&i_r(\bm{x},t) = \phi(\bm{x},t), &\bm{x}\in\partial V\label{eq:5}
\end{align}
with the, for now, unspecified time- and space-variant general boundary value $\phi$. At time $t = 0$, there are no particles in the sphere, so that $p(\bm{x},t = 0) = 0$.
\subsection{Vector Formulation}
\label{subsec:vec} 

For the following derivations, the IBVP in \eqref{eq:1}, \eqref{eq:2} is reformulated into a unifying vector formulation~\cite{rabenstein:ijc:2017}. Since most of the following derivations are performed in the continuous frequency domain\footnote{Please note that uppercase letters denote frequency domain variables and the complex frequency variable is denoted by $s$.}, the formulation is set up as 
\begin{align}
&\left[s\bm{C} - \Lp\right]\bm{Y}(\bm{x},s) = \bm{F}_\mathrm{e}(\bm{x},s), 
\label{eq:9}
\end{align}
where $\Lp = \bm{A} + \bm{\nabla}\bm{I} \in \mathbb{C}^{4\times 4}$ denotes a spatial differential operator. Matrix $\bm{C}\in \mathbb{C}^{4\times 4}$ is a capacitance matrix and \mbox{$\bm{A}\in \mathbb{C}^{4\times 4}$} is a matrix of damping parameters, depending on the diffusion coefficient $D$. 
The spatial derivatives in \eqref{eq:1}, \eqref{eq:2} are captured by the operator $\bm{\nabla}\bm{I}$. 
The matrices and operators in \eqref{eq:9} are derived by reformulation and Laplace $\mathcal{L}\{\cdot\}$ transformation of~\eqref{eq:1}, \eqref{eq:2} and are given as follows 
\begin{align}
&\bm{C} \!=\! \begin{bmatrix}
0 & 0 \\
1 & 0
\end{bmatrix}\!\!,\!\! \!\!
&\bm{A} \!= \!\begin{bmatrix}
\bm{0} & -\nicefrac{1}{D}\bm{I} \\
0 & \bm{0}
\end{bmatrix}\!\!, \!\!\!\!
&&\bm{\nabla}\bm{I} \!= \!\begin{bmatrix}
- \bm{\mathrm{grad}} & \!\!0 \\
0 & \!\!-\bm{\mathrm{div}}
\end{bmatrix}\!\!.\label{eq:8}
\end{align}
Moreover, vector $\bm{F}_\mathrm{e}\in \mathbb{C}^{4\times 1}$ contains the source term $f_\mathrm{e}$ from \eqref{eq:2}.  All physical quantities are collected in the vector 
\begin{align}
\bm{Y}(\bm{x},s) = \begin{bmatrix}
P(\bm{x},s) & 
\bm{I}(\bm{x},s)
\end{bmatrix}^\tra \in \mathbb{C}^{4\times 1}, \label{eq:7}
\end{align}
where $P = \mathcal{L}\{p\}$ and $\bm{I} = \mathcal{L}\{\bm{i}\}$ are the frequency domain equivalents of concentration $p$ and fluxes $\bm{i}$, respectively.

%

\section{Transfer Function Model}
\label{sec:model}
\begin{figure*}
	\centering
	\scalebox{0.9}{
		\centering
		\begin{tikzpicture}[auto, scale = 0.75, every node/.style={scale=0.75}, font=\large, node distance=2.5cm,>=latex', 
		rounded corners=2pt]
		
		\node[draw, dspsquare, minimum width=1cm, minimum height=1cm](s)at (8, 4) {
			\parbox{1cm}{
				\centering $s^{-1}$
		}};
		\node[draw, dspsquare, minimum width=1cm, minimum height=1cm](a)at (8, 2.5) {
			\parbox{1cm}{
				\centering $\bm{\mathcal{A}}_{\mathrm{S}1}$
		}};
		
		\node[draw, dspsquare, minimum width=2cm, minimum height=1cm](du)at (9, 1) {
			\parbox{2cm}{
				\centering $-\hat{\Bs}\hat{\Ks}$
		}};
		
		\node[draw,dspadder](add) at (6,4){};
		\node[draw,dspadder](add2) at (5,4){};
		\node[draw,dspadder](add3) at (19,4){};
		\node[draw,fill](p) at (10,4){};
		\node[draw,fill](p2) at (23,4){};
		\node[draw, dspsquare, minimum width=1.2cm, minimum height=1cm](c)at (12, 2.5) {
			\parbox{1.4cm}{
				\centering $\bm{\mathcal{C}}_{\mathrm{S}1}(\bm{x})$
		}};
		
		
		%
		\node[draw, dspsquare, minimum width=2cm, minimum height=1.2cm](trafo)at (15, 4) {
			\parbox{2cm}{
				\centering $\bm{T}_{\mathrm{S}1, \mathrm{S}2}$ 
		}};
		\node[draw, dspsquare, minimum width=1cm, minimum height=1cm](s2)at (21, 4) {
			\parbox{1cm}{
				\centering $s^{-1}$
		}};
		\node[draw, dspsquare, minimum width=1cm, minimum height=1cm](a2)at (21, 2.5) {
			\parbox{1cm}{
				\centering $\bm{\mathcal{A}}_\mathrm{S2}$
		}};
		
		\node[draw, dspsquare, minimum width=1.5cm, minimum height=1cm](c2)at (25, 4) {
			\parbox{1.5cm}{
				\centering $\bm{\mathcal{C}}_{\mathrm{S2}}(\bm{x})$
		}};
		
		\draw[dspflow,double] (3,4) node[dspnodeopen]{$\bar{\bm{F}}_{\mathrm{e},\mathrm{S}1}(s)$} to (add2) to (add);
		\draw[dspline,double] (add) to (s) to (p);
		\draw[dspline,double] (p) to (10,2.5) to (a);
		\draw[dspflow,double] (a) to (6,2.5); 
		\draw[dspline,double] (6,2.5) to (add);
		\draw[dspline,double] (p) to (12,4) to (c);
		\draw[dspline,double] (c) to (12,1) node[dspnodeopen]{}; 
		\node at (18,4.5) {
			\parbox{3cm}{
				\centering $\bar{\bm{\Phi}}_\mathrm{S2}(s)$
		}};
		\draw[dspflow,double] (11.9,4) -- (trafo);
		\draw[dspflow,double] (trafo) -- (add3);
		\node at (12,0.5) {
			\parbox{3cm}{
				\centering $\bm{Y}_{\mathrm{S}1}(\bm{x},s)$
		}};
		
		\draw[dspline,double] (add3) -- (s2) -- (p2) -- (c2);
		\draw[dspline,double] (p2) -- (23,2.5) -- (a2); 
		\draw[dspflow, double] (a2) -- (19,2.5);
		\draw[dspline,double] (19,2.5) -- (add3);  
		\draw[dspline,double] (c2) to (27,4) node[dspnodeopen]{}; 
		\node at (27,4.5) {
			\parbox{3cm}{
				\centering $\bm{Y}_{\mathrm{S2}}(\bm{x},s)$
		}};

		\node[draw,label=\Large$\gamma$](mult) at (7,1){};
		
		\draw[dspline,double] (11,4) -- (11,1) -- (du);
		\draw[dspflow,double] (du) -- (mult) -- (5,1) -- (5,1.5);
		\node[draw,dspnodeopen](du1) at (5,1.5){};
		\node[draw,dspnodeopen](du2) at (5,2.5){};
		\draw[dspflow,double] (du2) --node[midway,left]{$\bar{\bm{\Phi}}_{\mathrm{S}1}(s)$} (add2);
		\draw[dspline,double] (du1) -- (4.7,2.45);
		
		\node at (10,4.5) {
			\parbox{3cm}{
				\centering $\bar{\bm{Y}}_{\mathrm{S}1}(s)$
		}};
		\node at (23,4.5) {
			\parbox{3cm}{
				\centering $\bar{\bm{Y}}_\mathrm{S2}(s)$
		}};

		\node[draw, fill=gray!50, opacity=.2, dspsquare, minimum width=10.7cm, minimum height=5cm](c2)at (7.6, 2.5) {};
		\node[draw, fill=gray!50, opacity=.2, dspsquare, minimum width=11cm, minimum height=5cm](c2)at (22.5, 2.5) {};
		\node at (3.5,1.5) {
			\parbox{3cm}{
				\centering Sphere S1
		}};
		\node at (20,1.5) {
			\parbox{3cm}{
				\centering Sphere S2
		}};
		
		\path
		([shift={(-30\pgflinewidth,-10\pgflinewidth)}]current bounding box.south west)
		([shift={( 45\pgflinewidth, 10\pgflinewidth)}]current bounding box.north east);
		\end{tikzpicture}
	}
	\caption{\footnotesize Left: State-space description of the 3D diffusion process in the continuous frequency domain with state equation \eqref{eq:16} and output equation \eqref{eq:14} (switch open $\widehat{=}$ open loop system) and the attached feedback loop, according to \eqref{eq:32} (switch closed $\widehat{=}$ closed loop system). Complete figure: System model of two connected spheres according to Section~\ref{sec:inter}.}
	\label{fig:2}
\end{figure*}
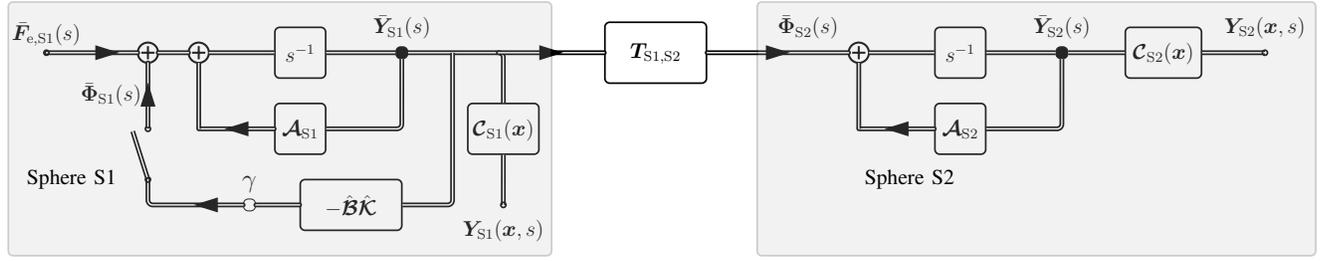
 The proposed modeling procedure is based on solving the IBVP presented in Section~\ref{subsec:ibvp} by modal expansion. The model is formulated in terms of transfer functions, i.e., in terms of a multidimensional SSD. 

For the modal expansion of the vector PDE~\eqref{eq:9}, an infinite set of bi-orthogonal eigenfunctions $\Kprim(\bm{x},\mu)\in \mathbb{C}^{4\times 1}$ and $\Kadj(\bm{x},\mu)\in \mathbb{C}^{4\times 1}$ is defined. The functions $\Kprim$ are the primal eigenfunctions and $\Kadj$ are their adjoints. The spectrum of the spatial differential operator $\Lp$ is of discrete nature and is defined by its eigenvalues $s_\mu$ \cite{churchill:1972}. The exact form of the eigenvalues is shown in Section~\ref{subsec:eig}, but index $\mu \in \mathbb{Z}$ is already introduced now to count the eigenvalues. The modal expansion of the PDE \eqref{eq:9} leads to a representation of the system in terms of a multidimensional SSD in a spatio-temporal transform domain \cite{Zwart:SCL:2004, rabenstein:ijc:2017}.

\subsection{Forward Transformation}

As the number of eigenfunctions and eigenvalues $s_\mu$ is infinite, we collect in vector $\bar{\bm{Y}}(s)$ all scalar transform domain representations $\bar{Y}(\mu,s)$ obtained by the transformation of vector $\bm{Y}(\bm{x},s)$. In order to match that formulation, the linear transformation operator $\tilde{\Cs}$ contains the corresponding infinitely many adjoint eigenfunctions $\Kadj(\bm{x},\mu)$, i.e., 
\begin{align}
&\bar{\bm{Y}}(s) \!= \!\left[\dots, \bar{Y}(\mu,s), \dots \right]^\tra\!\!\!, 
&\!\!\!\!\tilde{\Cs}(\bm{x}) \!= \!\left[ \dots, \Kadj(\bm{x},\mu), \dots \right]\!. 
\label{eq:11}
\end{align}
The entries of vector $\bar{\bm{Y}}(s)$ will act as the system states of the SSD in Section~\ref{subsec:ssd}. 
Exploiting \eqref{eq:11}, the forward transformation $\bm{\mathcal{T}}\left\lbrace\cdot\right\rbrace$ of the vector $\bm{Y}$ is formulated in terms of scalar products \cite{schaefer:dsp:2018}
\begin{align}
\bm{\mathcal{T}}\left\lbrace \bm{Y}(\bm{x},s) \right\rbrace = \bar{\bm{Y}}(s) = \linpr \bm{C} \bm{Y}(\bm{x},s), \tilde{\Cs}(\bm{x})\rinpr.
\label{eq:10}
\end{align}
Here $\linpr\bm{\cdot},\bm{\cdot}\rinpr$ denotes a vector of scalar products, where each scalar product is defined on the volume $V$ of the sphere
\vspace*{-1ex}
\begin{align}
\bar{Y}(\mu,s) &= \!\!\int_{V}\!\! \tilde{\bm{K}}^\her(\bm{x},\mu) \bm{C}\bm{Y}(\bm{x},s)\,\mathrm{d}\bm{x} = \linpr \bm{C}\bm{Y}, \Kadj \rinpr . 
\label{eq:12}
\end{align}
%
For the forward transformation in \eqref{eq:10}, the following differentiation theorem can be established \cite{rabenstein:ijc:2017} 
\begin{align}
&\linpr \Lp \bm{Y}(\bm{x},s), \tilde{\Cs}(\bm{x})\rinpr = \As \bar{\bm{Y}}(s) + \bar{\bm{\Phi}}(s),
\label{eq:13}
\end{align}
where $\As = \mathrm{diag}(\dots, \, s_\mu, \dots)$ is the diagonal operator of eigenvalues, acting as a state matrix. The vector $\bar{\bm{\Phi}}$ is the transform domain representation of the boundary values $\phi$ in \eqref{eq:5}, whose exact form is discussed in Section~\ref{sec:semi}.

\subsection{Inverse Transformation}

The inverse transformation $\bm{\mathcal{T}}^{-1}\!\!\left\lbrace \cdot \right\rbrace$ of \eqref{eq:10} exploits the discrete nature of the spectrum of the spatial differential operator $\Lp$, which allows a formulation in terms of a generalized Fourier series or in terms of scalar products \cite{churchill:1972, rabenstein:ijc:2017} 
\begin{align}
\begin{split}
\bm{\mathcal{T}}^{-1}\!\!\left\lbrace \bm{\bar{Y}}(s) \right\rbrace &= \sum_{\mu = 0}^{\infty} \frac{1}{N_{\mu}} \bar{Y}(\mu,s) \bm{K}(\bm{x},\mu)
= \Cs(\bm{x}) \bar{\bm{Y}}(s), 
\end{split}
\label{eq:14}
\end{align}
with the the inverse transformation operator $\Cs(\bm{x})$ containing all eigenfunctions $\Kprim(\bm{x},\mu)$ for the inverse transformation
\cite{rabenstein:ijc:2017}
\begin{align}
&\Cs(\bm{x}) = 
\begin{bmatrix}
\dots,\, \frac{1}{N_{\mu}}
\Kprim(\bm{x},\mu)
, \,\dots 
\end{bmatrix}.
\label{eq:15}
\end{align}
The scaling factor $N_\mu$ in \eqref{eq:14} and \eqref{eq:15} is derived from the bi-orthogonality between eigenfunctions $\Kprim$ and \mbox{$\Kadj$, i.e. $N_\mu = \langle\Kadj,\bm{C}\Kprim\rangle$} \cite{churchill:1972,rabenstein:ijc:2017}.
Together, \eqref{eq:10} and \eqref{eq:15} constitute a forward and inverse Sturm-Liouville transformation \cite{churchill:1972}.

\subsection{State-space Description}
\label{subsec:ssd}
Applying forward transformation \eqref{eq:10} to PDE \eqref{eq:9} and exploiting the differentiation theorem \eqref{eq:13} leads to a representation of the vector PDE in a spatio-temporal transform domain 
\begin{align}
s\bar{\bm{Y}}(s) = \As \bar{\bm{Y}}(s) + \bar{\bm{\Phi}}(s) +\bar{\bm{F}}_\mathrm{e}(s),
\label{eq:16} 
\end{align}
with the transform domain representation of the source functions $\bar{\bm{F}}_\mathrm{e}(s) = \linpr \bm{F}_\mathrm{e}(\bm{x},s), \tilde{\Cs}(\bm{x})\rinpr$. Hence, \eqref{eq:16} resembles the state equation of an SSD in the transform domain. Together with \eqref{eq:14}, both equations constitute an SSD of the solution $\bm{Y}(\bm{x},s)$ of the PDE \eqref{eq:9}, which is graphically shown on the left hand side of Fig.~\ref{fig:2} (switch open, neglecting subscript S$1$). Inspired by control theory, this system is referred to as the open loop system in the sequel.

\subsection{Eigenfunctions and Eigenvalues}
\label{subsec:eig}

The eigenfunctions $\Kprim$ and $\Kadj$ are derived from their dedicated eigenvalue problems \cite{rabenstein:ijc:2017,churchill:1972}. The detailed derivation is omitted here for brevity but it can be found in \cite{rabenstein:ijc:2017} for similar problems.
Solving the eigenvalue problem for the primal eigenfunctions $\Kprim$ leads to 
\begin{align}
\Kprim(\bm{x},\mu) \!= \!\begin{bmatrix}
j_n(k_{n,\nu}\, r)\, Y_n^m(\theta,\varphi) \\
-D\, k_{n,\nu}\, j_n'(k_{n,\nu}\, r)\, Y_n^m(\theta,\varphi)\\
D\,\frac{1}{r}\sin\theta \, j_n(k_{n,\nu}\, r)\, Y_n^{m'}(\theta,\varphi)\\
-D\, \mathrm{j}\frac{m}{r\sin\theta} \, j_n(k_{n,\nu}\, r)\, Y_n^m(\theta,\varphi)
\end{bmatrix}\!\!.\label{eq:19}
\end{align}
Here, function $j_n(\cdot)$ denotes the spherical Bessel function of order $n$ and $Y_n^m(\cdot)$ is the spherical harmonic function of order $n$ and degree $m = -n,\dots, n$. The first order derivatives of $j_n(\cdot)$ and $Y_n^m(\cdot)$ are denoted by $j_n'(\cdot)$ and $Y_n^{m'}(\cdot)$, respectively.
Index $\mu$, counting the eigenfunctions $\Kprim$, depends on indices $(n,\nu,m)$. Therefore, index $\mu$ can be seen as the mapping $(n, \nu, m) \rightarrow \mu$. 
The $k_{n,\nu}$ are the $\nu$-th real-valued zeros of 
\begin{align}
\frac{\partial}{\partial r}j_n(k_{n,\nu}\cdot r)\big\vert_{r = R_0} = 0, \label{eq:20}
\end{align}
which follows from the evaluation of boundary conditions for the eigenfunctions in relation to the Dirichlet boundary conditions in \eqref{eq:5} (see  \cite{rabenstein:ijc:2017,churchill:1972}). The real-valued eigenvalues $s_\mu$ are obtained during the derivation of eigenfunctions as follows
\begin{align}
s_\mu = -D\cdot k_{n,\nu}^2. \label{eq:21}
\end{align}
The eigenvalues $s_\mu$ do not depend on degree $m$, but there exist different eigenfunctions in \eqref{eq:19} for different values of $m$. Therefore, multiple eigenvalues occur, the number of which is increasing for increasing order $n$ \cite{ArfkenWeber:2001}.
An equation similar to \eqref{eq:19} can be established for the adjoint eigenfunctions $\Kadj$. 

\subsection{Source Function}
\label{subsec:source}
To model the distributed generation of particles, the source function $f_\mathrm{e}$ in \eqref{eq:2} has to be specified. In general, any arbitrary spatially and temporally distributed function can be used to model the particle generation in the sphere. In this paper, we consider the product of a temporal and a spatial source function, i.e., $f_\mathrm{e}(\bm{x},t) = f_\mathrm{t}(t)\cdot f_{\bm{x}}(\bm{x})$.
The temporal source function starts at $t = 0$ with a duration of $t_0$
\begin{align}
f_\mathrm{t}(t) = \begin{cases}
\frac{1}{2}\left(1 - \cos\left(\omega_0 t\right)\right) & 0 \le t \le t_0\\
0 & \text{else},
\end{cases}
\label{eq:e1}
\end{align}
with $\omega_0 = \frac{2\pi}{t_0}$. For multiple releases of particles of the form \eqref{eq:e1}, the function is repeated at different temporal positions. The spatial source function is centered at $r = 0$ and modeled by a raised cosine function
\begin{align}
f_{\bm{x}}(\bm{x}) = \begin{cases}
\frac{1}{2}\left(1 + \cos\left(\pi \frac{r}{r_0}\right) \right) & 0 \le r \le r_0\\
0 & \text{else}.
\end{cases}
\label{eq:e2}
\end{align}
The proposed source function allows the modeling of spatially and temporally non-instantaneous particle generation.

\subsection{Discrete-time Model}
\label{subsec:model:algo}

To numerically evaluate the dynamics of the diffusion process, the SSD \eqref{eq:14}, \eqref{eq:16} is transformed into the discrete-time domain by the application of an impulse-invariant transformation \cite{rabenstein:ijc:2017,schaefer:icc:2019} 
\begin{align}
\bar{\bm{y}}[k+1] &= \As^{\mathrm{d}}\bar{\bm{y}}[k] + T \bar{\bm{f}}_\mathrm{e}[k+1] + T \bar{\bm{\phi}}[k+1], \label{eq:36}\\
\bm{y}[\bm{x},k] &= \Cs(\bm{x}) \bar{\bm{y}}[k],
\label{eq:37}
\end{align}
where the discrete-time state matrix $\As^{\mathrm{d}} = \expE{\As T}$ is expressed in terms of the matrix exponential. The sampling interval is denoted by $T$, so that $t = kT$. The variables $\bar{\bm{f}}_\mathrm{e}$ and $\bar{\bm{\phi}}$ denote the discrete-time equivalents of $\bar{\bm{F}}_\mathrm{e}$ and $\bar{\bm{\Phi}}$ in \eqref{eq:16}, respectively.
Due to the infinite number of eigenvalues and eigenfunctions, the variables in \eqref{eq:36}, \eqref{eq:37} are of infinite size. For practical simulations, the number of eigenvalues $s_\mu$ and eigenfunctions $\Kprim$, $\Kadj$ is truncated to $\mu = 0, \dots, Q - 1$, which directly scales the accuracy of the algorithm \cite{schaefer:icc:2019}. By this truncation, the state matrix $\As^{\mathrm{d}}$ is of size $Q\times Q$, the transformation operators $\Cs$ become matrices of size $4\times Q$ and the state vector $\bar{\bm{y}}$ and transformed excitation $\bar{\bm{f}}_\mathrm{e}$ are of size $Q\times 1$. 

The derived model \eqref{eq:36}, \eqref{eq:37} of the spherical diffusion process with general boundary values $\phi$ is extended in Sections \ref{sec:semi} and \ref{sec:inter} to two application scenarios.

\section{Permeable Boundaries}
\label{sec:semi}
\begin{figure*}
	\begin{minipage}{0.49\linewidth}
		\vspace*{1.5ex}
		\centering
		\includegraphics[width=\linewidth]{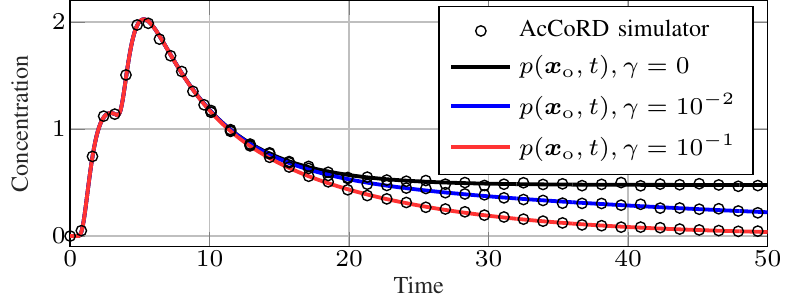}
	\end{minipage}
	\hfill
	\begin{minipage}{0.49\linewidth}
		\vspace*{-0.2ex}
		\centering
		\includegraphics[width=\linewidth]{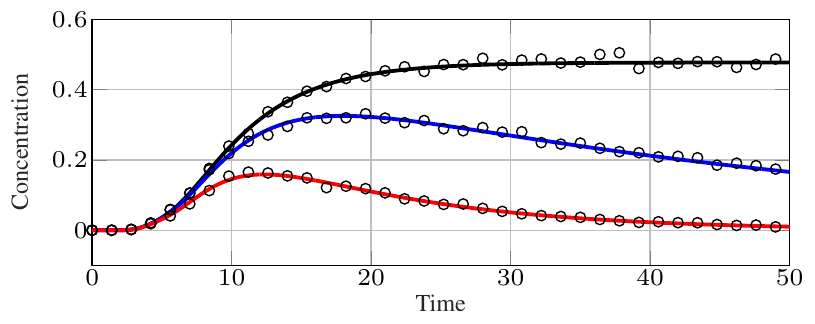}
	\end{minipage}
\vspace*{-1ex}
	\caption{Normalized concentration $p(\bm{x}_\mathrm{o},t)$ at $\bm{x}_\mathrm{o} = \left[r_1, \nicefrac{\pi}{3}, \nicefrac{\pi}{4}\right]$, with $r_1 = 0.4$ (left hand side) and $r_1 = 0.9$ (right hand side) over time for different values of $\gamma \in \{0, 10^{-2}, 10^{-1}\}$, derived by numerical evaluation of \eqref{eq:37}, \eqref{eq:35a} (black, blue, red) and with the AcCoRD particle simulator \cite{NOEL201744}.}
	\label{fig:4}
\end{figure*}
In the previous section, the model for particle diffusion in a bounded sphere was derived for the Dirichlet boundary conditions \eqref{eq:5} and the unspecified boundary values $\phi$. In this section, the SSD \eqref{eq:14}, \eqref{eq:16} is extended to two kinds of boundary conditions by the specification of the boundary values. For this purpose, the non-zero scalar values $\bar{\Phi}(\mu,s)$ of the transformed boundary term $\bar{\bm{\Phi}}(s)$ in \eqref{eq:13} have to be considered. The transformed boundary term is derived by evaluating the differentiation theorem \eqref{eq:13} and the corresponding Green's identities \cite{rabenstein:ijc:2017,churchill:1972}.
The individual elements $\bar{\Phi}(\mu,s)$ of $\bar{\bm{\Phi}}(s)$ are given in terms of the surface integral of the sphere                                            
\begin{align}
&\bar{\Phi}(\mu,s) = \int_{\partial V}
r^2\tilde{K}_4^*(\bm{x},\mu)\Phi(\bm{x},s)\sin\theta\,\mathrm{d}\varphi\,\mathrm{d}\theta, &\bm{x}\in\partial V, \label{eq:23}
\end{align}
with the surface of the sphere as defined in \eqref{eq:0}
and the fourth entry $\tilde{K}_4$ of the adjoint eigenfunctions $\Kadj$
\begin{align}
\tilde{K}_4^*(\bm{x},\mu) &= j_n(k_{n,\nu}\cdot r)\cdot Y_n^{m^*}(\theta,\varphi)\label{eq:24}.
\end{align}

\subsection{Non-permeable Boundary}
For a sphere with a non-permeable boundary, the particle flux in radial direction is zero at the boundary $r =R_0$ and all particles are reflected. Therefore, the boundary condition in \eqref{eq:5} is specified in the continuous frequency domain as follows
\begin{align}
&I_r(\bm{x},s) = \Phi(\bm{x},s) = 0, &\bm{x}\in\partial V. \label{eq:25}
\end{align}
Inserting \eqref{eq:25} into the transformed boundary values \eqref{eq:23} leads to $\bar{\bm{\Phi}}(s) = \bm{0}$ and the state equation \eqref{eq:16} simplifies to 
\begin{align}
s\bar{\bm{Y}}(s) = \As \bar{\bm{Y}}(s)  +\bar{\bm{F}}_\mathrm{e}(s).
\label{eq:26}
\end{align}
As no particles can leave the sphere, the injection of a finite number of particles leads to a saturation level $p_\infty$ of the concentration for $t \to \infty$.

\subsection{Semi-permeable Boundary}
\label{subsec:bc:semi}
In order to obtain a more realistic model (e.g., TX model in a MC system), the boundary of the sphere is assumed to be semi-permeable.
%
This allows the particles to leave the sphere, so that $p_\infty = 0$ for $t \to \infty$. The behavior is realized by impedance boundary conditions for the particle flux in radial direction
\begin{align}
&I_r(\bm{x},s)= \Phi(\bm{x},s) = \gamma\cdot P(\bm{x},s), &\bm{x}\in\partial V,
\label{eq:27}
\end{align}
where the permeability is controlled by the real-valued parameter $\gamma$. The impedance boundary conditions are incorporated into the model by the design of a feedback loop shifting the eigenvalues $s_\mu$ of the open loop system as defined in Section~\ref{sec:model} \cite{rabenstein:ijc:2017}.

The boundary value $\Phi$ in \eqref{eq:27} can be expressed by the output equation \eqref{eq:14} on the boundary $\partial V$
\begin{align}
\Phi(\bm{x},s) &= \gamma\cdot P(\bm{x},s) = \gamma\cdot\bm{c}_1^\tra(\bm{x})\bar{\bm{Y}}(s) \nonumber\\
&= \gamma \sum_{\hat{\mu} = 0}^{\infty} \frac{1}{N_{\hat{\mu}}}j_{\hat{n}}(k_{\hat{n}, \hat{\nu}}R_0)Y_{\hat{n}}^{\hat{m}}(\theta, \varphi)\bar{Y}(\hat{\mu},s),
\label{eq:30}
\end{align}
where $\bm{c}^\tra_1$ is the first row of the inverse transformation operator in \eqref{eq:15} and the index $\hat{\mu}$ is defined analogously to $\mu$, i.e., $(\hat{n}, \hat{\nu}, \hat{m}) \to \hat{\mu}$. 
Inserting \eqref{eq:24} and \eqref{eq:30} into \eqref{eq:23} leads to an expression for the scalar transformed boundary values $\bar{\Phi}(\mu,s)$ in terms of the system states 
\vspace*{-1ex}
\begin{align}
\bar{\Phi}(\mu,s) &= \gamma \, b(\mu) \sum_{\hat{\mu} = 0}^{\infty} \frac{1}{N_{\hat{\mu}}} j_{\hat{n}}(k_{\hat{n}, \hat{\nu}}R_0)\delta_{m,\hat{m}}\delta_{n,\hat{n}}\bar{Y}(\hat{\mu},s), \label{eq:31}
\end{align}
with $b(\mu) = R_0^2\cdot j_n(k_{n,\nu}R_0)$ and the Kronecker delta function $\delta$. Exploiting the formulation of the output equation in terms of the linear operators in \eqref{eq:14} leads to a representation of the transformed boundary term $\bar{\bm{\Phi}}(s)$ in \eqref{eq:23} in terms of the system states and a feedback matrix
\vspace*{-1ex}
\begin{align}
&\bar{\bm{\Phi}}(s) = -\gamma\cdot\hat{\Bs}\hat{\Ks}\bar{\bm{Y}}(s). 
\label{eq:32}
\end{align}
The exact form of the feedback matrix $\hat{\Bs}\hat{\Ks}$ is omitted for brevity. It can be obtained by arranging all $\mu = 0, \dots, Q -1$ values of $\bar{\Phi}(\mu,s)$ in \eqref{eq:31} into the vector $\bar{\bm{\Phi}}(s)$ and a reformulation in terms of state vector $\bar{\bm{Y}}$. 
Inserting \eqref{eq:32} into state equation \eqref{eq:16} leads to a modified state equation 
\vspace*{-0.5ex}
\begin{align}
s\bar{\bm{Y}}(s) = \left(\As - \gamma\cdot\hat{\Bs}\hat{\Ks}\right)\bar{\bm{Y}}(s)  +\bar{\bm{F}}_\mathrm{e}(s).
\label{eq:35} 
\end{align}
This modified state equation together with output equation \eqref{eq:14} leads to the SSD of the closed loop system (see Fig.~\ref{fig:2} left hand side, switch closed). Moreover, the modified state equation models the effect of the semi-permeable boundary \eqref{eq:27} of the sphere. The special case of a non-permeable (reflective) boundary \eqref{eq:25} is included for $\gamma = 0$. 
The modification of the state equation \eqref{eq:35} in the continuous frequency domain carries over into the discrete-time domain, which leads to a modified discrete-time state equation \eqref{eq:36}
\begin{align}
&\bar{\bm{y}}[k+1] = \As^{\mathrm{d}}_\mathrm{c}(\gamma)\,\bar{\bm{y}}[k] + T \bar{\bm{f}}_\mathrm{e}[k+1], \label{eq:35a}
\end{align}
with closed loop state matrix 
\begin{align}
\As^{\mathrm{d}}_\mathrm{c}(\gamma) = \expE{(\As - \gamma\hat{\Bs}\hat{\Ks})T}
. \label{eq:35b}
\end{align}
We note that the formulation in \eqref{eq:35a} allows parameter $\gamma$ to be varied over time during the numerical evaluation, i.e., \mbox{$\gamma\to\gamma[k]$}. This facilitates the modeling of boundaries with time-variant permeability. 
%

\subsection{Numerical Evaluation}
\label{sec:semi_num_eval}
In the following, we numerically evaluate the proposed model for a semi-permeable sphere, which may act as a spherical TX in a MC system. 
The scenario follows Fig.~\ref{fig:1}, where the emission of particles from the sphere is controlled by its permeability.
We use a sphere with a normalized radius of $R_0 = 1$ and a normalized diffusion coefficient of $D = 10^{-2}$. Through the spatially and temporally distributed source function $f_\mathrm{e}$ from Section~\ref{subsec:source}, the model allows to consider non-instantaneous particle production. 
It can be used to model realistic generation of signaling particles in a spherical TX of a MC system. 
%
Moreover, the particles are released at the center of the sphere at $0.25\si{s}$ and $3\si{s}$, respectively.
A single release is non-uniformly spatially distributed in a sphere with radius $r_0 = 0.1 R_0$ and it takes $t_0 = 0.1\si{\second}$ until all particles are released.
The results of the numerical evaluation are shown in Fig.~\ref{fig:4} for the normalized concentration $p(\bm{x}_\mathrm{o},t)$, using the closed loop SSD \eqref{eq:37}, \eqref{eq:35a} with $Q = 240$ eigenvalues\footnote{The number of $Q = 240$ eigenvalues provides a good trade-off between accuracy and evaluation time.}. The concentration dynamics are observed at different observation points given by $\bm{x}_\mathrm{o} = \left[r_1, \nicefrac{\pi}{3}, \nicefrac{\pi}{4}\right]$, with $r_1 = 0.9\,R_0$ and $r_1 = 0.4\,R_0$, and for different permeabilities $\gamma \in \{0, 10^{-2}, 10^{-1}\}$.
We observe that for a reflective boundary ($\gamma = 0$) the concentration saturates to a non-zero final value for $t \to \infty$. By 
increasing the permeability of the sphere ($\gamma \neq 0$) all particles can eventually leave the sphere and, thus, the concentration will be zero for $t \to \infty$. Moreover,  
depending on the observation point it is possible 
to observe the impact of the two 
separate releases of particles. 
On the left hand side ($r_1 = 0.4\,R_0$) of Fig.~\ref{fig:4}, it is still possible to identify both releases in the step-wise increase of the concentration between $0\,\si{\second}$ and $7\,\si{\second}$.  
However, on the right hand side ($r_1 = 0.9\,R_0$) it is impossible to determine the exact number of individual particle releases. Due to the diffusion in the sphere, both particle releases are merged into a single slope in the concentration progression.
%
We verified our results with particle-based simulations using the AcCoRD simulator~\cite{NOEL201744}. We observe that the numerical evaluation and the simulation results match perfectly. However, the proposed model outperforms the simulator in terms of run-time. The simulator needed approx. $15\,\si{\minute}$ to obtain an individual concentration curve in Fig.~\ref{fig:4}, while the numerical evaluation of the proposed model required only $15\,\si{\second}$.
%

\begin{figure}[t]
	\centering
	\includegraphics[width=\linewidth]{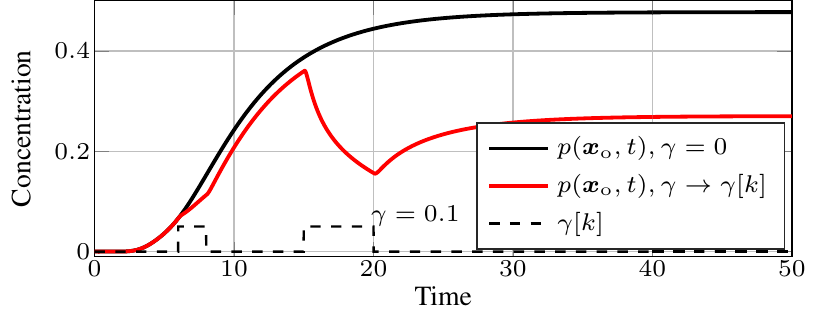}
	\vspace*{-4ex}
	\caption{Normalized concentration $p(\bm{x}_\mathrm{o},t)$ at $\bm{x}_\mathrm{o} = \left[r_1, \nicefrac{\pi}{3}, \nicefrac{\pi}{4}\right]$, with \mbox{$r_1 = 0.9\,R_0$} simulated with the model \eqref{eq:37}, \eqref{eq:35a} for $\gamma = 0$ and time-variant $\gamma \to \gamma[k]$. The variation of permeability $\gamma[k]$ is also shown.}
	\label{fig:5}
	\vspace*{-2ex}
\end{figure}

In Fig.~\ref{fig:5}, we consider the case of a time-variant permeability, i.e., $\gamma\to\gamma[k]$. Compared to the results in Fig.~\ref{fig:4} the 
particle release from the boundary of the sphere is not constant, but is controlled by a time-variant permeability. The temporal progression of $\gamma[k]$ is illustrated by the dashed curve in Fig.~\ref{fig:5} and toggles between $0$ and $0.1$. We clearly observe the influence of the permeability on the concentration. 
%
The results in Fig.~\ref{fig:5} show that the proposed model is able to realize a time-variant permeability. This adds a further degree of freedom to the design of spherical TX with the proposed model. The particle release from the TX is specified on the one hand by radius $R_0$ of the sphere, the diffusion coefficient $D$, and the particle generation function $f_\mathrm{e}$. On the other hand, the release of the particles into the environment can be regulated by a time-variant permeability. 

\section{Interconnected Spheres}
\label{sec:inter}

%

In addition to modeling a (time-variant) semi-permeable boundary, the general model in Section~\ref{sec:model} can also be used as the basis for modeling the interconnection of two spherical objects (see Fig.~\ref{fig:3}).
In the following, we consider a simple scenario from inter-cell communication, where two cells are connected by a semi-permeable membrane and ion channels.
For simplicity the following assumptions are made:
\begin{itemize}
	\item Spheres S1 and S2 have the same geometrical dimensions $R_0$ and the same diffusion coefficient $D$.
	\item Spheres S1 and S2 are both reflective except for a semi-permeable membrane having the area \mbox{$\partial V_\mathrm{c} \coloneqq \{\bm{x} \big\vert r = R_0, \varphi \in [-\pi, \pi], \theta\in [0, \theta_0] \}$} on the surface of S1 and S2, respectively. The permeability of S1 and S2 is controlled by $\gamma_{\mathrm{S}1}$ and $\gamma_{\mathrm{S}2}$, respectively.
	\item The permeable areas of S1 and S2 are connected by linear delay-free ion channels with diode behavior, so that particles can only pass from sphere S1 to S2 (see Fig.~\ref{fig:3}, red area on the right hand side).
\end{itemize}
%
\subsection{Interconnection Matrix}
The boundary values for the partially permeable boundary of S1 are defined by a spatial truncation of \eqref{eq:27} as follows 
\begin{align}
\Phi_{\mathrm{S1}}(\bm{x},s) = 
\begin{cases}
\gamma_{\mathrm{S}1}\cdot P_{\mathrm{S1}}(\bm{x},s) & \bm{x} \in \partial V_\mathrm{c}\\
0 & \bm{x}\notin\partial V_\mathrm{c}
\end{cases},\label{eq:39}
\end{align}
%
where the value of the permeability $\gamma_{\mathrm{S}1}$ depends on the number and characteristics of the ion channels.
The subscript S1 in \eqref{eq:39} indicates values belonging to sphere S1. 
The particle concentration $P_{\mathrm{S}1}$ is expressed in terms of the system states of S1 by reducing  \eqref{eq:14} to 
\begin{align}
&P_{\mathrm{S}1}(\bm{x},s) = \bm{c}^\tra_{\mathrm{S}1}(\bm{x})\bar{\bm{Y}}_{\mathrm{S}1}(s), &\bm{x} \in {\partial V_\mathrm{c}}, \label{eq:41}
\end{align}
where $\bm{c}^\tra_{\mathrm{S}1}$ is the first row of the transformation operator in \eqref{eq:15} of S1.
With the boundary values in \eqref{eq:39}, S1 is permeable for $\bm{x} \in \partial V_\mathrm{c}$ and reflective for $\bm{x} \notin \partial V_\mathrm{c}$. To realize the desired connection of S1 and S2, the boundary values for sphere S2 are defined analogously to Kirchhoff's current law. The particle flux for $\bm{x} \in \partial V_\mathrm{c}$ into sphere S2 is given by the particle flux $\Phi_{\mathrm{S1}}(\bm{x},s)$ from S1. Therefore, the boundary values for S2 are defined by 
\begin{align}
\Phi_{\mathrm{S2}}(\bm{x},s) = 
\begin{cases}
-\gamma_{\mathrm{S}2}\Phi_{\mathrm{S1}}(\bm{x},s) &\bm{x} \in \partial V_\mathrm{c}\\
0 & \bm{x} \notin \partial V_\mathrm{c}
\end{cases}\label{eq:40}.
\end{align}
Subsequently, the transformed boundary values $\bar{\bm{\Phi}}_{\mathrm{S}2}$ serving as input of sphere S2 (see Fig.~\ref{fig:2}) can be derived by a transformation of the system states $\bar{\bm{Y}}_{\mathrm{S}1}$ of S1 into the scalar entries $\bar{\Phi}_{\mathrm{S}2}(\mu,s)$ of $\bar{\bm{\Phi}}_{\mathrm{S}2}$. This transformation is performed by redefining \eqref{eq:23} for sphere S2 on the surface part $\bm{x} \in \partial V_\mathrm{c}$ 
\begin{align}
\bar{\Phi}_{\mathrm{S}2}(\mu,s) = \int_{\partial V_\mathrm{c}}
r^2\tilde{K}_{4, \mathrm{S}2}^*(\bm{x},\mu)\Phi_{\mathrm{S}2}(\bm{x},s)\,\mathrm{d}\bm{x}. \label{eq:42}
\end{align}
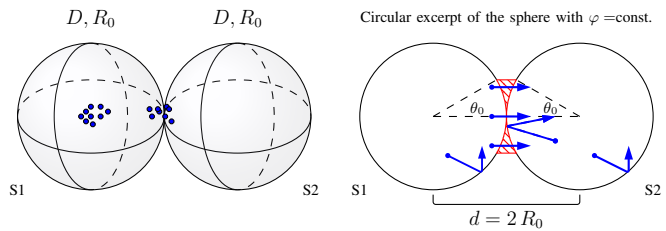
\begin{figure}[t]
	\centering
	\begin{tikzpicture}[auto, scale = 0.65, every node/.style={scale=0.65}, font=\LARGE, node distance=2.5cm,>=latex', 
	rounded corners=1pt,domain=0:10]
	
	\draw (-1.5,2.5) arc (180:360:1.5cm and 0.75cm);
	\draw[dashed] (-1.5,2.5) arc (180:0:1.5cm and 0.75cm);
	\draw (0,4) arc (90:270:0.75cm and 1.5cm);
	\draw[dashed] (0,4) arc (90:-90:0.75cm and 1.5cm);
	\draw (0,2.5) circle (1.5cm);
	\shade[ball color=blue!10!white,opacity=0.10] (0,2.5) circle (1.5cm);
	
	\draw (1.5,2.5) arc (180:360:1.5cm and 0.75cm);
	\draw[dashed] (1.5,2.5) arc (180:0:1.5cm and 0.75cm);
	\draw (3,4) arc (90:270:0.75cm and 1.5cm);
	\draw[dashed] (3,4) arc (90:-90:0.75cm and 1.5cm);
	\draw (3,2.5) circle (1.5cm);
	\shade[ball color=blue!10!white,opacity=0.10] (3,2.5) circle (1.5cm);
	
	
	\node at (0,4.5){
		\parbox{2cm}{
			\centering \large$D, R_0$
		}
	};
	\node at (3,4.5){
		\parbox{2cm}{
			\centering \large$D, R_0$
		}
	};

	%
	\draw[fill=blue] (0,2.5) circle (0.05cm);
	\draw[fill=blue] (0,2.7) circle (0.05cm);
	\draw[fill=blue] (0.2,2.5) circle (0.05cm);
	\draw[fill=blue] (-0.2,2.5) circle (0.05cm);
	\draw[fill=blue] (-0.1,2.6) circle (0.05cm);
	\draw[fill=blue] (0.04,2.34) circle (0.05cm);
	\draw[fill=blue] (-0.1,2.4) circle (0.05cm);
	\draw[fill=blue] (0.2,2.7) circle (0.05cm);
	\draw[fill=blue] (0.35,2.6) circle (0.05cm);
	
	\draw[fill=blue] (1.4,2.6) circle (0.05cm);
	\draw[fill=blue] (1.52,2.5) circle (0.05cm);
	\draw[fill=blue] (1.55,2.7) circle (0.05cm);
	\draw[fill=blue] (1.37,2.65) circle (0.05cm);		
	\draw[fill=blue] (1.6,2.65) circle (0.05cm);
	\draw[fill=blue] (1.65,2.4) circle (0.05cm);
	\draw[fill=blue] (1.2,2.65) circle (0.05cm);
	\draw[fill=blue] (1.25,2.5) circle (0.05cm);		
	
	
	\draw[black] (7,0.9) -- (7,0.7) -- node[midway,below]{\large$d = 2\,R_0$} (10,0.7) -- (10,0.9);
	

	\draw (7,2.5) circle (1.5cm);
	\draw (10,2.5) circle (1.5cm);
	
	\draw[black,dashed] 
	(10,2.5) 
	-- ++(150:1.5)  
	arc[start angle=150, end angle=210, radius=1.5cm];
	\draw[white,thick] 
	(10,2.5) 
	++(150:1.5) node[coordinate](a){} 
	arc[start angle=150, end angle=210, radius=1.5cm] node[coordinate](b){};
	
	\draw[black,dashed] 
	(7,2.5) 
	-- ++(30:1.5)  
	arc[start angle=30, end angle=-30, radius=1.5cm];
	\draw[white,thick] 
	(7,2.5) 
	++(30:1.5)  node[coordinate](c){}
	arc[start angle=30, end angle=-30, radius=1.5cm] node[coordinate](d){};
	
	\path[pattern=north west lines, pattern color=red,draw=red,rounded corners=0pt] (a) arc[start angle=150, end angle=210, radius=1.5cm] --(d) arc[start angle=-30, end angle=30, radius=1.5cm];
	\draw[red] (c) -- (a);
	\draw[dashed](7,2.5) -- node[xshift=1ex, yshift=-0.5ex,above,midway]{\small$\theta_0$} (8.5,2.5);
	\draw[dashed](8.5,2.5) -- node[xshift=1ex, yshift=-0.5ex,above,midway]{\small$\theta_0$} (10,2.5);
	
	
	\draw[dspconn,blue,thick] (8.2,2.5)node[thick,circle,fill=blue,inner sep=0pt,minimum size=3pt]{} -- (9,2.5);
	\draw[dspconn,blue,thick] (8.2,1.9) node[thick,circle,fill=blue,inner sep=0pt,minimum size=3pt]{}-- (9,1.9);
	\draw[dspconn,blue,thick] (8.2,3.1) node[thick,circle,fill=blue,inner sep=0pt,minimum size=3pt]{}-- (9,3.1);	
	
	\draw[dspconn,blue]  (7.3,1.7) node[thick,circle,fill=blue,inner sep=0pt,minimum size=3pt]{} -- (8,1.35) -- (8,1.9);
	\draw[dspconn,blue,thick] (10.3,1.7) node[thick,circle,fill=blue,inner sep=0pt,minimum size=3pt]{}-- (11,1.35) -- (11,1.9);
	
	\draw[dspconn,blue](9.5,2) node[thick,circle,fill=blue,inner sep=0pt,minimum size=3pt]{}-- (8.5,2.3) -- (9.5,2.5);	
	
	\draw[dspconn,dashed,blue](9.5,2) node[thick,circle,fill=blue,inner sep=0pt,minimum size=3pt]{}-- (8.5,2.3) -- (9.5,2.5); 
	\node[inner sep=0pt] (lab) at (8.5,4.5){\small Circular excerpt of the sphere with $\varphi=$const.};
	
	\node[inner sep=0pt] (lab) at (-1.5,1){\small S1};
	\node[inner sep=0pt] (lab) at (5.5,1){\small S1};
	
	\node[inner sep=0pt] (lab) at (4.5,1){\small S2};	
	\node[inner sep=0pt] (lab) at (11.5,1){\small S2};
	\end{tikzpicture}
	\vspace*{-2ex}
	\caption{Left: Conceptual system model of two identical bounded spheres. Particles are released in sphere $\mathrm{S}1$. Right: At the intersection, particles can penetrate the permeable boundary of $\mathrm{S}1$ to enter $\mathrm{S}2$ via delay-free ion channels (red). Particles can traverse from S1 to S2 but not backwards via the ion channels and are reflected on the non-permeable boundaries (blue trajectories in the right sphere).}
	\label{fig:3}
	\vspace*{-2ex}
\end{figure}
\noindent Inserting \eqref{eq:40} into \eqref{eq:42} leads to the desired representation of the transformed boundary values $\bar{\Phi}_{\mathrm{S}2}$ of S2 in terms of the system states $\bar{\bm{Y}}_{\mathrm{S}1}$ of S1
\vspace*{-0.7ex}
\begin{align}
\bar{\Phi}_{\mathrm{S}2}(\mu,s) = -\gamma_{\mathrm{S}1}\gamma_{\mathrm{S}2} R_0^2\int_{\partial V_\mathrm{c}} \!\tilde{K}_{4, \mathrm{S}2}^*(\bm{x},\mu) \bm{c}^\tra_{\mathrm{S}1}(\bm{x})\,\mathrm{d}\bm{x}\, \bar{\bm{Y}}_{\mathrm{S}1}(s). \label{eq:43}
\end{align}
Analogously to \eqref{eq:32}, the transformed boundary values $\bar{\bm{\Phi}}_{\mathrm{S}2}$ of S2 can be expressed in terms of a connection matrix $\bm{T}_{\mathrm{S}1, \mathrm{S}2}$ that transforms the system states $\bar{\bm{Y}}_{\mathrm{S}1}$ of S1 as follows
\vspace*{-0.7ex}
\begin{align}
\bar{\bm{\Phi}}_{\mathrm{S}2}(s) = \bm{T}_{\mathrm{S}1, \mathrm{S}2}(\gamma_{\mathrm{S}1},\gamma_{\mathrm{S}2})\bar{\bm{Y}}_{\mathrm{S}1}. \label{eq:43a}
\end{align}
The exact form of connection matrix $\bm{T}_{\mathrm{S}1, \mathrm{S}2}$ is omitted for brevity. It can be derived by evaluating the integral in \eqref{eq:43} and arranging all $\mu = 1, \dots, Q-1$ values of $\bar{\Phi}_{\mathrm{S}2}(\mu,s)$ into the vector $\bar{\bm{\Phi}}_{\mathrm{S}2}(s)$.

Hence, the complete system of the connected spheres S1 and S2 can be formulated in terms of an SSD, which is illustrated in~Fig.~\ref{fig:2} (closed switch).

\subsection{Numerical Evaluation}

In the following, we numerically evaluate the dynamics of the concentration in two interconnected spheres S1 and S2 based on the the previously introduced model. Similar to Section~\ref{sec:semi_num_eval}, the radius of the sphere is 
$R_0=1$, the diffusion coefficient is given by $D=10^{-2}$. Particles are generated by the source function $f_\mathrm{e}$ from Section~\ref{subsec:source}. 
They are released at the center of S1 at $0.25\,\si{s}$ and $3\,\si{s}$, respectively. A single release is non-uniformly spatially distributed in a sphere with radius $r_0 = 0.4\,R_0$ and it takes $t_0= 0.4\,\si{\second}$ until all particles are released. The angle defining the permeable area $\partial V_\mathrm{c}$  of the spheres is given by $\theta_0 = \nicefrac{\pi}{4}$. The permeability in the spheres S1 and S2 is $\gamma_{\mathrm{S}1} = 0.1$ and $\gamma_{\mathrm{S}2} = 1$ (fully permeable), respectively. Moreover, the concentration is observed in S1 at $\bm{x}_\mathrm{o,S1} = \left[R_0, \nicefrac{\pi}{2}, 0\right]$ and in S2 at $\bm{x}_\mathrm{o,S2} = \left[r_2, \nicefrac{\pi}{2}, 0\right]$, with $r_2 = R_0$ and $r_2=0.1R_0$. 

\begin{figure}[t]
	\centering
	\includegraphics[width=\linewidth]{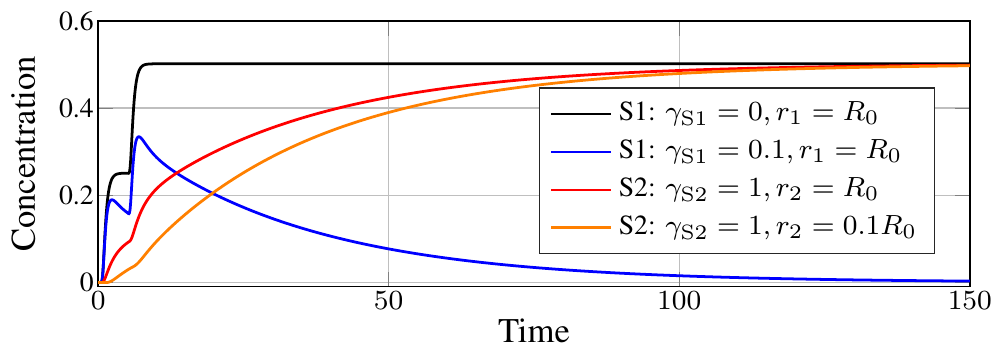}
	\vspace*{-4ex}
	\caption{Particle concentration in  S1 and S2 over time. Concentration in S1 for reflective boundary ($\gamma_{\mathrm{S}1}= 0$) and with partially permeable boundary at $\bm{x} \in \partial V_\mathrm{c}$ at $r_1 = R_0$. Particle concentration in S2 at $\bm{x}_\mathrm{o,S2}$ for $r_2 = R_0$ and $r_2 = 0.1 R_0$. }
	\label{fig:6}
	\vspace*{-3ex}
\end{figure}
The results of the numerical evaluation are shown in Fig~\ref{fig:6}. 
We observe that the concentration in S2 ($\gamma_\text{S1}= 0.1$,$\gamma_\text{S2}= 1$)
saturates to the concentration for $\gamma_\text{S1}= 0$ in S1. 
%
%
This indicates that all particles traverse from S1 to S2 via the ion channels for $t\to\infty$. Moreover, we observe that the concentrations in S1 ($\gamma_\text{S1} = 0.1$) at $r_1 = R_0$ and S2 at $r_2 = R_0$ show a complementary behavior. 
The concentration in S2  at $r_2 = 0.1\,R_0$ shows the dynamics of the particles further away from the boundary. 
Inspecting the corresponding orange curve in Fig.~\ref{fig:6}, the abrupt change in the slope 
at $10\,\si{\second}$ reveals that 
the influence of the second particle release in S1 at $3\,\si{\second}$ is still observed in S2. 
%

The results in Fig.~\ref{fig:6} and the preceding derivations show that, with the proposed transfer function approach  
the concentration dynamics in cascaded spherical objects, e.g. simplified cell models, can be conveniently modeled by the interconnection of individual spheres. Due to this block based modeling approach, the scenario can be extended to longer cell chains and more complex interconnection topologies.  

\section{Conclusions}
\label{sec:conc}
\vspace*{-0.2ex}
We derived a transfer function model for particle diffusion in a sphere with general boundary behavior. The model was derived by the application of functional transformations to an initial-boundary value problem and was formulated in terms of a state-space description. Based on this general model we studied two application scenarios. 
A sphere with a (time-variant) semi-permeable boundary was considered as a flexible model for a spherical transmitter, whose characteristics can be adjusted by its dimension, diffusion coefficient, and a time-variant permeability.
%
Furthermore, the interconnection of two spheres by a semi-permeable membrane and delay-free ion channels was considered, which can serve as a basis for modeling inter-cell communication and Ca\textsuperscript{$2+$} signaling.
We numerically evaluated the proposed model and verified the results by particle-based simulations. 

\bibliographystyle{IEEEtran}
{\footnotesize
	\bibliography{./../Bib/IEEEAbrv,%
		./../Bib/icc20_final}

\end{document}